\documentclass[twocolumn,aps,prl,showpacs]{revtex4}
\usepackage[dvips]{graphicx}
\usepackage{bm}
\begin{document}
%\draft
%\twocolumn[\hsize\textwidth\columnwidth\hsize
%\csname @twocolumnfalse\endcsname
\title{Effect of dipolar interactions on the magnetization of a cubic array of nanomagnets}
\author{Marisol Alc{\'a}ntara Ortigoza}
\email{alcantar@phys.ksu.edu}
\author{Richard A. Klemm}
\email{klemm@phys.ksu.edu}  \author{Talat S.
Rahman}\email{rahman@phys.ksu.edu} \affiliation{Department of Physics,
Kansas State University, Manhattan, KS 66506 USA}
%xxx.lanl.cond-mat/0501006
\date{\today}
\begin{abstract}
We investigated the effect of intermolecular dipolar interactions
on an ensemble of 100 3D-systems of $5\times5\times4$ nanomagnets,
each with spin $S = 5$, arranged in a cubic lattice. We employed
the Landau-Lifshitz-Gilbert  equation to solve for the
magnetization curves for several values of the damping constant,
the induction sweep rate, the lattice constant, the temperature,
and the magnetic anisotropy. We find that the smaller the damping
constant, the stronger the maximum induction required to produce
hysteresis. The shape of the hysteresis loops also depends on the
damping constant. We find further that the system magnetizes and
demagnetizes at decreasing magnetic field strengths with
decreasing sweep rates, resulting in smaller hysteresis loops.
Variations of the lattice constant within realistic values (1.5 nm
- 2.5 nm) show that the dipolar interaction plays an important
role in the magnetic hysteresis by controlling the relaxation
process.  The temperature dependencies  of the damping constant
and of the magnetization are presented and discussed with regard
to recent experimental data on nanomagnets. Magnetic anisotropy
enhances the size of the hysteresis loops for external fields
parallel to the anisotropy axis, but decreases it for
perpendicular external fields. Finally, we reproduce and test a
previously reported magnetization curve for a 2D-system [M. Kayali
and W. Saslow, Phys. Rev. B {\bf 70}, 174404 (2004)]. We show that
its hysteretic behavior is only weakly dependent on the shape
anisotropy field and the sweep rate, but depends sensitively upon
the dipolar interactions.  Although in 3D systems, dipole-dipole
interactions generally diminish the hysteresis, in two-dimensional
systems, they strongly enhance it. For both square two-dimensional
and rectangular three-dimensional lattices with ${\bm
B}||(\hat{\bm x}+\hat{\bm y})$, dipole-dipole interactions can
cause large jumps in the magnetization.

\end{abstract}
 \vskip0pt \maketitle

\section{Introduction}
The need of smaller memory storage
devices,\cite{newns,kf1,chen,smyth,fk1,kf2,fk2,fk3,koehler,kw,bb1,bb2,deutsch,zhang}
the interest in developing quantum computing,\cite{loss} and the
hope for understanding the relationship between the macroscopic
and microscopic magnetic behaviors has led   intense  research
into the properties of nanoscale
magnets.\cite{newns,kf1,chen,smyth,fk1,kf2,fk2,fk3,koehler,kw,bb1,bb2,deutsch,zhang,loss,luis,sangregorio,hill,aubin,bokacheva,ohm,perenboom,wernsdorfer1,ek,tiron,wernsdorfer2,wernsdorfer3,vanslageren,friedman,thomas,wernsdorfer4,wernsdorfer5}
Many issues still remain unclear and serious problems must be
overcome in order for them to be technologically useful. Prominent
among these is the loss of memory during magnetic relaxation.

Ferromagnetic nanodots are complex systems consisting of up to
hundreds of magnetic atoms within a single dot.\cite{fk1,bb1,bb2}
In this case, interparticle interactions along with anisotropy
effects dominate the dynamics of the systems, and  control the
magnetization processes.\cite{fk3} Moreover, since interdot
exchange interactions are negligibly small, the dynamics of the
ferromagnetic nanodot arrangements are strongly influenced by
dipolar interdot interactions.\cite{deutsch,zhang}

Single molecule magnets (SMM's) consist of clusters of only a few
magnetic ions, and are thus among the smallest and simplest
nanomagnets, but are also well-characterized systems exhibiting
magnetic hysteresis.\cite{wernsdorfer3} In SMM's, the one-body
tunnel picture of the magnetization mostly explains this
phenomenon in the sense that the sequence of discrete steps in
those curves provides evidence for  resonant coherent quantum
tunneling.\cite{vanslageren,friedman,thomas} Nevertheless, this
one-body tunnel model neglects intermolecular interactions, and is
not always sufficient to explain the measured tunnel
transitions.\cite{wernsdorfer4,wernsdorfer5}  A close examination
of the magnetization curves reveals fine structures which cannot
be explained by that model. Wernsdorfer {\it et al.} suggest that
these additional steps are due to collective quantum processes,
called spin-spin cross relaxation (SSCR), involving pairs of SMM's
which are coupled by dipolar and/or exchange
interactions.\cite{wernsdorfer4,wernsdorfer5} If dipolar and/or
exchange interactions cooperate in the relaxation process, then
one might hope to be able to better control such
 loss of magnetic memory.

 Analyzing the relaxation
of the magnetization is difficult for both SMM's and ferromagnetic
nanodots.  Besides dipolar interactions, many other factors may be
involved in such processes. Geometric features, such as the shape
and volume of the magnets, as well as the type of lattice on which
they are placed, can directly influence the anisotropy barriers
and
 the easy axis directions. In the case of SMM's, a quantum
treatment has to be considered to show that resonant tunneling of
the magnetization results in the discrete steps appearing in the
low temperature $T$ magnetization curves. Although in many SMM's
the intercluster exchange interactions are negligible, as for
ferromagnetic nanodots, in other SMM's, such interactions are
comparable in strength to the dipolar
interactions.\cite{wernsdorfer5} Besides the quadratic Heisenberg
and quadratic anisotropic intramolecular exchange interactions,
some SMM's are thought to contain intramolecular interactions of
the Dzyaloshinskii-Moriya type.\cite{deraedt} Additional higher
order,
 anisotropic  spin exchange interactions further complicate the problem.
 Therefore, by studying
models that deal with each one of these factors separately, one
hopes to simplify the problem, to build up gradually a more
realistic system, and at the same time, to elucidate how each  of
these factors contributes to the magnetization process.

 With regard to SMM's, there have been recent approaches to the
quantum dynamics of the low-$T$
relaxation.\cite{sangregorio,ps,fa1,ts,fa2,santini,furukawa}
Prokof'ev and Stamp  assumed a single relaxation mode,\cite{ps} in
which the dipolar and hyperfine fields are frozen unless  an SMM
flips its spin.  Then by assuming  the effective field around each
SMM is that of randomly placed dipoles, they obtained  an
expression for the low-$T$ decrease proportional to $t^p$ of the
magnetization of each SMM from its fully magnetized
state,\cite{ps,anderson1,mw} where  $p\approx0.5-0.7$, but $p$
might be as large as 0.7.\cite{ps,ts,fa1,fa2}. This procedure was
restricted to very small deviations of the magnetization from its
saturated value, so it is not useful for studying the central
portion of the hysteresis curves, for which the magnetization can
be small. Moreover, as first argued for ferromagnets by
Anderson,\cite{anderson2}  the spin-spin and spin-lattice
relaxation times can be very different, so that such simple
behavior is not expected. In fact, experiments on SMM's have shown
that an exponential relaxation of the magnetization is consistent
with the data,\cite{santini,furukawa} so that as a minimum, one
requires two distinct relaxation times for SMM's, which could be
very different from one another.\cite{anderson2}

 The most commonly studied model of spin  dynamics containing
two distinct relaxation parameters is the Landau-Lifshitz-Gilbert
(LLG) equation.\cite{lifshitz,huang} Using the LLG equation,
Kayali and Saslow (KS) investigated the hysteresis curves at $T=0$
for two-dimensional (2D) square arrays of 4 to 169 ferromagnetic
nanodots subject to dipole-dipole interactions and a magnetic
field applied in various directions within the array's $xy$
plane.\cite{saslow} They included anisotropy effects via an
effective field proportional to the $z$-component of each dot's
dipole moment. Earlier studies of square planar lattices of 9 to
36 ferromagnetic dots were made by
  Stamps and Camley.\cite{camley}  In addition,   Zhang
  and Fredkin (ZF) studied the LLG model to obtain the zero-field
  time decay  of the easy-axis magnetization of a three-dimensional (3D) cubic
  lattice of 12$\times12\times12$ Stoner-Wohlfarth particles
  interacting with each other via dipole-dipole
  interactions.\cite{zhang} Since the size (or radius) of the
  Stoner-Wohlfarth particles was taken to be much less than the
  lattice constant,  they could be treated as point-like
  magnetic moments, the classical analogue of SMM's.

Here we study only the effects of the intermolecular dipole-dipole
interactions upon the magnetization curves for an ensemble of
$N_c=100$  3D cubic crystals each containing $N=5\times5\times4$
nanomagnets, all with the same magnetic moment. As in the ZF model
of Stoner-Wohlfarth particles, we take the lattice parameter to be
much greater than the  nanomagnet size  or radius.   Except when a
strong anisotropy field is present, we assume that there is no
long-range order in the $T$ regime of interest, so that in the
absence of an external magnetic field, the magnetization of each
nanomagnet crystal is essentially zero. We note that long-range
ordering was claimed to exist in such a system with Ising spin
anisotropy.\cite{FA1,evangelisti} In our studies with a strong
anisotropy field ${\bm H}_A$, hysteresis curves exhibiting a
substantial zero-field magnetization were obtained  for the
applied magnetic induction ${\bm B}\>||\>{\bm H}_A$ after the
system had been fully magnetized by ${\bm B}$.  The strength of
the dipole interactions is  primarily determined by the lattice
constant, $a$, which we vary from 1.25 nm to 2.5 nm. The dynamics
of each nanomagnet are assumed to be given by the LLG equation,
which includes precession and damping relaxation processes, the
damping  coefficient $\alpha$ of which can also depend upon
$T$.\cite{FR,WB} Then, the magnetic moment ${\bm M}_i^c$ of the
$i$th  nanomagnet within the $c$th crystal of our ensemble obeys
\begin{eqnarray}
\frac{d{\bm M}_i^c}{dt}&\equiv&\gamma{\bm M}_i^c\times{\bm
B}_i^{c,\rm eff}-\frac{\alpha}{M_s}{\bm M}_i^c\times\bigl({\bm
M}_i^c\times{\bm B}_i^{c,\rm eff}\bigr),\\
 {\bm B}_i^{c,\rm eff}&=&{\bm
B}+\bigl({\bm B}_i^c\bigr)_{\rm dip},\label{beff}
\end{eqnarray}
where $\gamma=g\mu_B$ is the gyromagnetic ratio,  $M_s=g\mu_BS$ is
the magnetic moment of an individual nanomagnet, and $\bigl({\bm
B}_i^c\bigr)_{\rm dip}$ is the contribution to the effective
magnetic induction ${\bm B}_i^{c,\rm eff}$ at the $i$th
 nanomagnet within the $c$th crystal arising from dipole-dipole
interactions between it and the other  nanomagnets within the same
crystal,
\begin{eqnarray}
\bigl({\bm B}_i^c\bigr)_{\rm
dip}&=&\frac{\mu_0}{4\pi}\sum_{j}{'}\>\frac{3({\bm M}_j^c\cdot{\bm
r}_{ij}){\bm r}_{ij}-r_{ij}^2{\bm
M}_j^c}{r_{ij}^5},\label{bdipole}
\end{eqnarray}
where the prime indicates that the $j=i$ term is omitted. The
second term of Eq. (1), the damping term, was first introduced by
Landau and Lifshitz \cite{lifshitz} and later by Gilbert to give a
phenomenological description of the relaxation of the
magnetization.  They did not derive it from first principles due
to the enormous complexity of summarizing all of the relaxation
processes into a single term. As noted above, when ferromagnetic
interactions are present, $\alpha/\gamma$ is generally expected to
be $\ll1$.\cite{anderson2} By extending to electronic spin systems
the Wangsness-Bloch model of nuclear spin magnetic relaxation by
magnetic dipole coupling to a heat bath,\cite{WB} Fredkin and Ron
showed that the damping term could be derived for large spin
values and $\kappa=\hbar\gamma H/k_BT\ll1$, where $\hbar$ and
$k_B$ are Planck's constant divided by $2\pi$ and Boltzmanns'
constant, respectively, and in our case $H=B_i^{c,\rm
eff}$.\cite{FR} To the extent that electric quadrupole
interactions could be neglected, $\alpha$ varies inversely with
$T$ for $\kappa\ll1$, but depends upon $\kappa$
otherwise.\cite{FR}

 More recently,  a different derivation of the Gilbert damping
term  was derived from a spin Hamiltonian containing the
interaction between the spin and the radiation field, which is
induced by the precessing magnetization itself.\cite{ho1,ho2}  In
that case, no explicit $T$ dependence of $\alpha$ was given.  We
remark that rather complex explicit expressions for $\alpha$  for
the different system of local spin moments arising from $p-d$
kinetic-exchange coupling of the itinerant-spin subsystem in
ferromagnetic semiconductor alloys
 have been given
recently.\cite{sinova}  In any event, the damping coefficient
$\alpha$ at some $T$  value must be determined experimentally for
each system.

In order to study the magnetization of ferromagnetic dots, KS used
an extremely large value for the damping coefficient,
$\alpha/\gamma = 0.6$,  a huge sweep rate, $\frac{\Delta B}{\Delta
t}\sim3000$ T/s, and a small maximum external induction $B_{\rm
max}=2\mu_0M_s$.  In our studies of  nanomagnet arrays, we used
values of $\alpha/\gamma$ that  varied from these values to values
12 orders of magnitude smaller.  Depending upon the $\alpha$
values, we also  varied the sweep rate $\frac{\Delta B}{\Delta t}$
 from those values to the the much smaller $\sim10^{-3}$ T/s,
and  varied $B_{\rm max}$  from much larger values (2 T),
comparable to those reported in SMM
experiments,\cite{wernsdorfer1,wernsdorfer2}  to those used by KS.
Similarly, the lattice constants reported in the present work are
in accordance with the   near neighbor separation in the most
extensively studied SMM crystals.   Quantum processes within the
individual SMM's will be treated in a separate
publication.\cite{marisol2}

The present paper is organized follows. In Sec. II, we present our
model system and a brief description of the overall calculation
procedure that we followed. In Sec. III, we solve the LLG for each
 nanomagnet subject to both the external and the combined dipolar
inductions. In Sec. IV, we present  and discuss our main results
for the magnetization curves, which are evaluated at various
values of the sweep rate, $T$, $a$, the anisotropy field, and
$\alpha/\gamma$.  When spin anisotropy is present, we  study the
cases ${\bm B}\>||\>{\bm H}_A$ and ${\bm B}\!\perp\!{\bm H}_A$.
 In Sec. V, we reproduce one of the KS magnetization curves
for square 2D lattices, and vary some of their parameters to show
that the results are almost independent of the sweep rate over the
range $\sim300-6000$ T/s. Analogously, we show that the anisotropy
field does not affect significantly the magnetization curves until
its magnitude is comparable to $B_{\rm max}/\mu_0$. By varying the
lattice constant,  we also show that the results of KS are very
sensitive to the strength of the dipolar fields, which mainly
govern the behavior of the magnetization of such systems. Finally,
in Sec. VI,  we summarize our main conclusions.

\section{Model System}
In the present work we consider an ensemble  of $N_c=100$ cubic
crystals (or configurations), each configuration consisting of
$N=5\times5\times4=100$ nanomagnets,  each with ground state spin
$S = 5$, which interact with one another only via dipolar
interactions. Each  of the $N_c$ system configurations
$c=1,\ldots,N_c$ is constructed to have a starting total magnetic
moment ${\bm M}_c \approx 0$ at ${\bm B} = 0$. The hysteresis
curves are obtained for each configuration,  and these are then
averaged over the $N_c$ configurations. One then obtains the
magnetization $\overrightarrow{\cal M}(B)$ curves, where
$\overrightarrow{\cal M}=\langle {\bm M}_c\rangle_c/V$ is the
configuration averaged magnetization, $V$ is the crystal volume,
and $B=|{\bm B}|$.

\subsection{Ensemble of random spin configurations}
In order to proceed, we first  find a large number $N_c$ of random
spin configurations $c$  of $N=100$ spins, such that for each
configuration, ${\bm M}_c/M_s\approx 0$ at ${\bm B} = 0$ and as
$T\rightarrow\infty$, where  the total magnetic moment
\begin{eqnarray}
{\bm M}_c(t,{\bm B})&=&\sum_{i=1}^{N=100}{\bm M}_i^c(t,{\bm B}).
\end{eqnarray}
  At the start of the
iteration, we take $t=0$, ${\bm B} = 0$, and $T\rightarrow\infty$
in the absence of the dipole-dipole (or any other inter-spin)
interactions for configuration $c$.  Then we select those
configurations for which $|{\bm M}_c|/M_s \le 0.1$,  which we deem
sufficiently close to ${\bm M}_c\approx0$. Our resulting
magnetization curves are based  upon the average over $N_c = 100$
configurations, each one containing $N=100$  similarly chosen
nanomagnets.

 We reiterate that $N$ is the number of nanomagnets in each
configuration, and $N_c$ is the number of configurations studied.
Although we have  chosen both of these numbers to be 100 in order
to obtain reliable statistics, $N$ and $N_c$ have completely
different meanings. Finding many  $(N_c$) configurations,  each of
which has  an almost vanishing  initial magnetization  consumes a
significant amount of computer time, especially if the number $N$
of nanomagnets per configuration is not very large. However,
choosing a rather small number $N$ of nanomagnets reduces the time
required to calculate the dipolar field at each nanomagnet due to
all  of its neighbors,  which must be calculated at each
integration time step of the LLG equation, offsetting the large
amount of computer time required to set up $N_c$ initially
nearly-nonmagnetic configurations.

\subsection{Evolution of the magnetization versus  field curves}
In this model one increases the external magnetic induction ${\bm
B}$ in discrete steps $\Delta {\bm B}$, until ${\bm B} = {\bm
B}_{\rm max}$, where $B_{\rm max}=|{\bm B}_{\rm max}|$ has to be
large enough  to align every  nanomagnet in its direction. How
large $B_{\rm max}$ has to be generally depends upon $T$, the
field sweep rate $\frac{\Delta B}{\Delta t}\equiv\frac{|\Delta
{\bm B}|}{\Delta t}$, the lattice parameter $a$, and the crystal
structure.\cite{wernsdorfer3} In addition, the steps $|\Delta {\bm
B}|$ must be small enough to give rise to numerically smooth
${\cal M}(B)$ curves. We therefore set $|\Delta {\bm B}| = B_{\rm
max}/N_B$, where the number of steps $N_B \gg 1$ should be on the
order of $10^3$.  After each magnetic step, one allows each of the
nanomagnets to  relax for a fixed amount of time $\Delta t$, which
is chosen to be sufficiently small  that the nanomagnets do not
reach equilibrium. Otherwise,
 in the absence of a sufficiently strong anisotropy field, no
hysteresis would result.

First, we choose one of our configurations $c$ (e. g., $c=1$) and
set the moments of the  nanomagnets equal to their values in this
 initially nonmagnetic configuration, $\{{\bm M}_i^{c=1} (t=0,
{\bm B} = 0)\}_{i=1,\ldots,N}$. That is, just after we turn on the
magnetic induction in the $x$ direction by the amount ${\bm B} =
\Delta {\bm B}$, the  nanomagnets have not yet precessed from
their initial configuration. Then, we calculate the effective
magnetic induction ${\bm B}_i^{c=1,\rm eff}$  at each of the
$i=1,\ldots, N$  nanomagnets for $c=1$. To do so, we must
calculate the dipolar induction in Eq. (3) due to the presence of
all the other  nanomagnets.

 Then, we let each of the  nanomagnets evolve
in the presence of its effective magnetic induction for a  chosen
fixed time interval $\Delta t$. To do this accurately, we break
$\Delta t$ up into $N_t$ intervals $dt = \Delta t/N_t$. Obviously,
this is extremely time consuming, because it is necessary to
recalculate the effective induction at each  nanomagnet after each
time-integration step  of width $dt$. Once the whole set of
moments $\{{\bm M}_i^1 (t= \Delta t, {\bm B} = \Delta{\bm
B})\}_{i=1,\ldots, N}$ is obtained, we  proceed to calculate the
configuration magnetic moment, ${\bm M}_{c=1} (\Delta t, \Delta
{\bm B})$, for this choice of  fixed sweep rate, $\Delta B/\Delta
t$, from
\begin{eqnarray}
{\bm M}_{c}(t,{\bm B})&=&\frac{\sum_{i=1}^N{\bm M}_i^{c}(t,{\bm
B})\exp[-\beta{\cal
H}_i^{c}(t,{\bm B})]}{\sum_{i=1}^N\exp[-\beta{\cal H}_i^{c}(t,{\bm B})]},\label{statistics}\\
 {\cal H}_i^c(t,{\bm B})&=&-{\bm B}_i^{c,\rm eff}(t,{\bm B})\cdot{\bm
M}_i^{c}(t,{\bm B}),\label{mc}
\end{eqnarray}
 by setting $c=1$, $t=\Delta t$ and ${\bm
B}=\Delta{\bm B}$, where ${\bm B}_i^{c,\rm eff}(t,{\bm B})$ is
given from Eqs. (\ref{beff}) and (\ref{bdipole}), $\beta=1/k_BT$,
and $k_B$ is Boltzmann's constant.  Since $\bigl({\bm
B}_i^{c}\bigr)_{\rm dip}$ as given by Eq. (\ref{bdipole}) in ${\bm
B}_i^{c,\rm eff}(t,{\bm B})$ contains a self-fieldless single sum,
there is no site overcounting in Eq. (\ref{mc}).

We are interested in ${\bm M}_c({\bm B},\frac{\Delta B}{\Delta
t})$, a thermodynamic quantity that does not explicitly depend
upon $t$. Although the ${\bm M}_i^c$ and ${\bm B}_i^{c, \rm eff}$
for each nanomagnet are evaluated at each time step $dt$, the
statistical weighting in Eqs. (\ref{statistics}) and (\ref{mc}) is
only evaluated at the
 end of each  fixed interval $\Delta t$, which has
 a one-to-one correspondence with $\Delta B$. Thus, this
 single-configuration average can be directly compared to those in
 different configurations after the same number of intervals.
 Moreover, since $\Delta t=\Delta B\left(\frac{\Delta B}{\Delta
 t}\right)^{-1}$, ${\bm M}_c(t,{\bm B})$ for our purpose can be written  as
 ${\bm M}_c(\frac{{\bm B}}{\Delta B/\Delta t},{\bm B})$, which is effectively a function of ${\bm B}$ and $\frac{\Delta B}{\Delta t}$.

 Next, we increase the external magnetic induction by another equal step $\Delta{\bm
 B}$,
  and let the  nanomagnets precess during another equal time
interval, $\Delta t$, under the action of the new effective induction.
 We
 continued increasing ${\bm B}$ in this equal step fashion a
total of $N_B$ times, until ${\bm B} = {\bm B}_{\rm max}$. At this
point, the incremental induction direction is reversed, setting
${\bm B} = {\bm B}_{\rm max} - \Delta{\bm B}$ for the same time
interval $\Delta t$,  repeating the procedure  $2N_B$ times, until
${\bm B} = - {\bm B}_{\rm max}$. After that, we reverse the
incremental induction direction once again, setting ${\bm B} = -
{\bm B}_{\rm max} + \Delta{\bm B}$  for the same time interval
$\Delta t$, etc., and continue $N_B$ times until ${\bm B} = 0$ is
reached, or until the  configuration magnetization hysteretic loop
(if it exists) is closed. Then, one repeats the entire procedure
above described for each of the other $N_c-1$ configurations $c=2,
\ldots, N_c$,  keeping the time intervals $\Delta t$ and the
subintervals $dt$ constant for each step in each configuration.
Once all  of the calculations for each of the $N_c$ configuration
are finished, we average the results over the $N_c$
configurations, obtaining,
\begin{eqnarray}
\langle{\bm M}_c\Bigl(\frac{{\bf B}}{\Delta B/\Delta t},{\bm
B}\Bigr)\rangle_c&=&\frac{1}{N_c}\sum_{c=1}^{N_c}{\bm
M}_c\Bigl(\frac{{\bm B}}{\Delta B/\Delta t},{\bm B}\Bigr).
\end{eqnarray}
Then, the magnetization $\overrightarrow{\cal M}$
 is easily calculated. Having
tabulated $\overrightarrow{\cal M}$ for every external magnetic
induction step with fixed
 $\frac{\Delta B}{\Delta t}$,  $T$, $a$,  $M_s$,  $\alpha$, and $B_{\rm max}$, we
generate the magnetization curve ${\cal M}(B)$ for this set of
parameters.

\subsection{Variation of experimental parameters}
 Unlike the parameters such as $B_{\rm max}$ and $dt$, which are
details of the theoretical calculation, other parameters can in
principle be varied in experiments in a variety of materials.
 Using the same initial dipole configurations we
repeat the whole procedure with different values of $\alpha$,
$\frac{\Delta B}{\Delta t}$,  $T$, and  $a$.  The only parameters
that can be experimentally varied in studies on a particular
sample are $\frac{\Delta B}{\Delta t}$ and $T$,  since the other
parameters are fixed. Nevertheless, the possibility of setting the
 nanomagnets further apart by varying the composition of the
non-magnetic ligand groups  in SMM's, for example, justifies the
study of the variation of
 $a$.   Also, given that the damping term
appearing in the LLG equation is phenomenological, and that in
most cases $\alpha$ should be determined experimentally, we have
also examined its variation.   We note that $\alpha$ is expected
to depend inversely upon $T$, unless $T$ is sufficiently low that
thermal processes  no longer dominate the relaxation.\cite{zhang}
We keep $M_s$ fixed.

\section{Integration of the LLG equation for one nanomagnet}

The magnetic moment of each nanomagnet is obtained by numerically
integrating the LLG equation. The time-evolution of one nanomagnet
must be determined synchronously with all its neighbors in order
to calculate the dipolar induction acting on each of them at a
given time. To solve the LLG equation for the $i$th nanomagnet in
the $c$th crystal, we first rotate its coordinates at each time
integration step such that ${\bm B}_i^{c,\rm eff}(t)\>||\>\hat{\bm
z}(t)$.  We then solve the resulting differential equations for
either the coordinate spherical angles $\theta(t), \phi(t)$, or
the components of ${\bm M}_i^c(t)$, as shown in Appendix A. The
quantity relevant to each  spherical angle or component of ${\bm
M}_i^c(t)$ is $\int_{t_0}^td\tau|{\bm B}_i^{c,\rm eff}(\tau)|,$
which explicitly involves the past history of $|{\bm B}_i^{c,\rm
eff}(t)|$.   In order to decrease the computation time, we
approximate this integral for small time integration steps
$dt=t-t_0\ll t_0$,
\begin{eqnarray}
\int_{t_0}^td\tau|{\bm B}_i^{c,\rm eff}(\tau)|&\approx&|{\bm
B}_i^{c,\rm eff}(t_0)|dt.\label{memory}
\end{eqnarray}

 In order to assure numerical accuracy of our results for the
greatly different experimental parameters studied, we had to make
appropriate choices for the numerical parameters used in the
calculations,  as discussed in the Appendix.  Generally,
calculations with slow sweep rates require correspondingly small
$\alpha/\gamma$ values. For the calculations leading to the
results presented in Figs. 1, 2, and 6-9, we take the numerical
parameters $dt = 1\times10^{-4}$ s, $B_{\rm max} = 2.0$ T, $N_t =
1000$, and $N_B = 500, 1000$ and 4000, respectively, for the
different sweep rates studied.  For the calculations presented in
Figs.  3-5, we take $dt=6\times10^{-12}$ s, $B_{\rm max}=22.5$ mT,
$N_t=1000$, and $N_B=1250$.

\section{Results and discussion}
\subsection{Effects of damping and maximum induction values on the hysteresis}
  We first neglect any spin anisotropy effects.  In Fig. 1,  we  plotted the average over
$N_c=100$ configurations of the normalized magnetization at the
lattice constant $a$ = 1.5 nm, sweep rate $\frac{\Delta B}{\Delta
t} = 0.005$ T/s, and temperature $T=0.7$ K for the four weak
damping rates $\alpha/\gamma=3\times10^{-n}$, where $n=10, 11,
12,$ and 13.  These results appear respectively from left to right
(right to left) in the upper (lower) part of Fig. 1. The
magnetization curves show hysteresis for all four of these
$\alpha$ values.  For the smallest $\alpha$ value we studied,
$\alpha/\gamma = 3.0 \times 10^{-13}$, the hysteresis only occurs
for  external induction magnitudes exceeding 3.0 T,  observed by
setting $B_{\rm max}$ above that value, which is well beyond the
domain pictured in Fig. 1. We also note that in Fig. 1, the
central regions for $|\langle M\rangle/(NM_s)|<0.8$ are
non-hysteretic. For each of these four parameter value choices,
the initial curve describing the first increase of the average
magnetization from essentially 0 to its saturation value is
indistinguishable from subsequent similar curves obtained after
completing the full hysteresis paths. Hence, in this case, the
main consequence of the choice of $N_c=100$ configurations is the
improvement in the statistics, reducing the noise that remains
most evident in the curves corresponding to the smallest $\alpha$
values.

   From the inset to Fig. 1, we see that although
the height (in $<M>/(NM_s)$)  of the hysteretic region decreases
with decreasing $\alpha$, the width (in $B$) of the hysteretic
region increases faster with decreasing $\alpha$, so that the
overall area of the hysteretic region increases with decreasing
$\alpha$. From a computational standpoint, for the parameter
values studied in Fig. 1, the  smaller the value of $\alpha$, the
 larger the required value of $B_{\rm max}$  to produce
hysteresis.  We also noticed that in these magnetization curves,
the hysteresis sets in at the point of an abrupt change in slope
in the initial curve, which describes the first increase of the
average magnetization from 0 to its saturation value. Moreover, we
conclude that  $B_{\rm max}$ must be chosen to guarantee that the
system reaches saturation at ${B}\le B_{\rm max}$, because of the
different nature of the hysteresis present in each curve.  For
example, in Fig. 1 the hysteresis can occur only after saturation,
but with smaller $a$ values, if the system has not saturated by
$B=B_{\rm max}$, then the magnetization will keep increasing for a
number of subsequent $\Delta {\bm B}$ steps, even though the
direction of $\Delta {\bm B}$ (but not of ${\bm B}$) has been
reversed.
\begin{figure}
\includegraphics[width=0.45\textwidth]{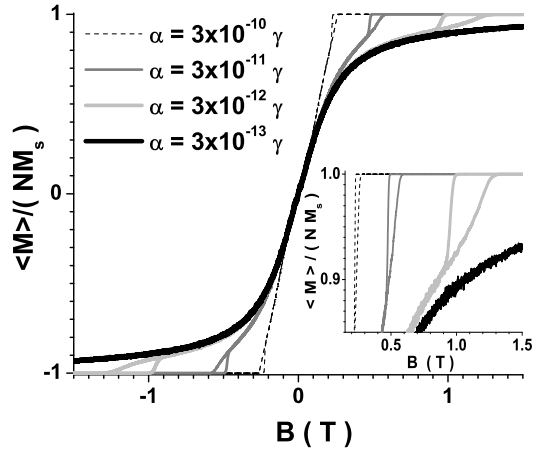}
\caption{Magnetization curves for  $N_c=100$, $a$ = 1.5 nm,
$\frac{\Delta B}{\Delta t} = 0.005$  From left to right for ${\cal
M}>0$, $\alpha/\gamma = 3\times10^{-10}$ (dashed),
$3\times10^{-11}$ (thin dark grey), $3\times10^{-12}$ (light
grey), $3\times10^{-13}$ (thick black). The inset highlights the
hysteretic region of the first three of these curves.}
\label{fig1}
\end{figure}

\subsection{Effect of temperature on the hysteresis}

\subsubsection{Temperature-independent $\alpha$}
\begin{figure}
\includegraphics[width=0.45\textwidth]{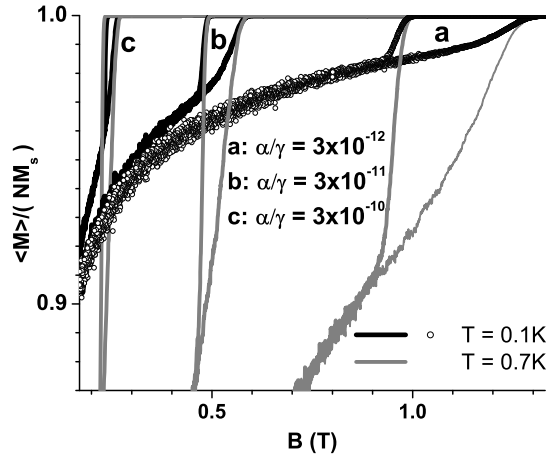} \caption{Shown is the upper hysteretic
region of the normalized magnetization curves at  $T=0.7$ K (grey)
and $T=0.1$ K (black, circles). The $T$-independent damping
constants $\alpha/\gamma$ are $3\times10^{-12}$ (a),
$3\times10^{-11}$ (b), and $3\times10^{-10}$ (c). The other
parameters are the same as in Fig. 1.} \label{fig2}
\end{figure}

 We first investigate the role of temperature that arises only
from the statistics, Eq. (\ref{statistics}), and present our
results for a $T$-independent $\alpha$ in Fig. 2.  In this figure,
we have replotted the inset of Fig. 1, excluding the curve for
$\alpha/\gamma=3\times10^{-13}$, for which the hysteresis occurred
for $B$ too large to display on the same plot.  Otherwise, the
parameters are the same as in Fig. 1, except that we have compared
our results (grey curves) for $T=0.7$ K shown in Fig. 1 with those
(black curves and circles) for $T=0.1$ K.  Since the evolution of
the magnetization with ${\bm B}$ in this model is independent of
$T$, we note from Fig. 2 that the departures of the magnetization
curves from the points of saturation are the same at both $T$
values, so that the widths (in $B$) of the hysteretic regions are
nearly the same. However, the height in $< M>/(NM_s)$ of each
hysteretic region decreases strongly with decreasing $T$, so that
the overall area of each hysteretic region decreases with
decreasing $T$. This particular result is in strong contrast to
the existing experimental results on SMM's. Nevertheless, our
results are reasonable from the point of view of the LLG equation
and the way $T$ enters our calculation.  We reiterate that we have
so far neglected quantum and spin anisotropy effects, the latter
of which will be discussed in the following.

 We remark  that  in Fig. 2, $T$ only enters into the
equations of motion when the average magnetic moment is evaluated
 from Eq. (\ref{statistics}).   As for the Brillouin function
that describes the magnetization of a paramagnet in the absence of
the dipole interactions, the initial slope of the magnetization at
low $B$ increases as  $T$ is lowered.  This increases the
alignment of the moment of each nanomagnet at decreasing $T$, so
that the dipole-dipole interactions tend to be maximized,
enhancing the effect.

\subsubsection{Temperature-dependent $\alpha$}

 We now consider the effect of the temperature dependence of the
damping constant $\alpha$ upon the magnetic hysteresis, focussing
upon the case of correspondingly fixed very high sweep and damping
rates. We assume that our choice of spin value, $S=5$ for each
nanomagnet, satisfies $S\gg1$. In this limit, Fredkin and Ron
showed that the damping of the nuclear spin precession by magnetic
dipole coupling to a heat bath, as derived under the assumption of
spin-orbit factorization by Wangsness and Bloch, could be readily
extended to the spins in magnetic systems.\cite{FR,WB} For
$S\gg1$, they found
\begin{eqnarray}
\alpha(T)/\gamma&\approx&\frac{T_0}{T},\label{alphaofT}
 \end{eqnarray}
  where
 $T_0=2\hbar\Phi_{11}^1(1-e^{-\kappa})S^2/{k_B\kappa}$,\cite{FR}
 and $\Phi_{11}^1$ is a  rate constant (with units of $s^{-1}$),
 the expression for which is a complicated orbital integral arising from
 the interaction of the local spin with its surrounding molecular
 electronic orbitals  in second-order perturbation theory,\cite{WB} and $\kappa=\hbar\gamma
 { B}^{\rm eff}/(k_BT)$.  For $\kappa\ll1$, $T_0\rightarrow
 2\hbar\Phi_{11}^1S^2/k_B$, which can be taken to be independent
 of $T$ and ${ B}^{\rm eff}$, so that $\alpha\propto T^{-1}$, but
 for $\kappa\gg1$, $\alpha\propto 1/{ B}^{\rm eff}$, which
 would completely change its effect.  Here we only consider the
 case $\kappa\ll1$, for which Eq. (\ref{alphaofT}) holds for
 constant $T_0$.  We note that, as in Figs. 1 and 2, $T$ also affects the results
 for the magnetization from the statistics, Eq. (\ref{statistics}).

  In Fig. 3, we have shown our results, averaged over $N_c=100$
configurations, of  the normalized
 magnetization as a function of $B$ in mT, for $a=1.5$ nm,
 $\frac{\Delta B}{\Delta t}=3000$ T/s,  $\alpha(T)/\gamma=T_0/T$, $T_0=0.3 K$, and
 $T=5$ K.   For the calculations presented in this figure, we used the numerical parameters
 $dt=6\times10^{-12}$ s, $B_{\rm max}=22.5$ mT, $N_t=1000$, and $N_B=1250.$
 Note that although $a$ has the same value as in Figs. 1
 and 2, the sweep and damping ($\alpha(T)/\gamma=0.06$) rates are six and at least eight
 orders of magnitude larger than in those figures.  For these
 parameters, there are three regions of hysteresis in the pictured
 magnetization curve.  The left inset is an enlargement of the
 upper hysteretic region, the mirror image of which occurs in the
 lower region of the pictured magnetization  curve.  In contrast to the behavior shown in Figs. 1 and 2,
  at the top of the upper hysteretic region, the
 magnetization does not rise abruptly to its saturation value, but
 first goes through an extended non-hysteretic region.  In
 addition, there is a central hysteretic region, an enlargement of
 which is shown in the right inset, along with an enlargement of
 the same central hysteretic region obtained at $T=0.25$ K with
 the same set of parameters.  We note that at $T=5$ K, the onset
 magnetization averaged over $N_c=100$ configurations,  pictured by the thin curve
 in the lower portion of the right inset, does not coincide with the thick curve
 corresponding to the central hysteresis loop region obtained subsequently to the attainment of the saturation value by the magnetization.
 In addition, we note that the thick central hysteresis loop exhibits
 reproducible oscillations  with $B$-independent frequency $f$ at $T=5$K, which oscillations have
 disappeared at the lower $T=0.25$ K  value, for which $\alpha(T)/\gamma=1.2$, pictured in the upper
 portion of the right inset.

\begin{figure}
\includegraphics[width=0.45\textwidth]{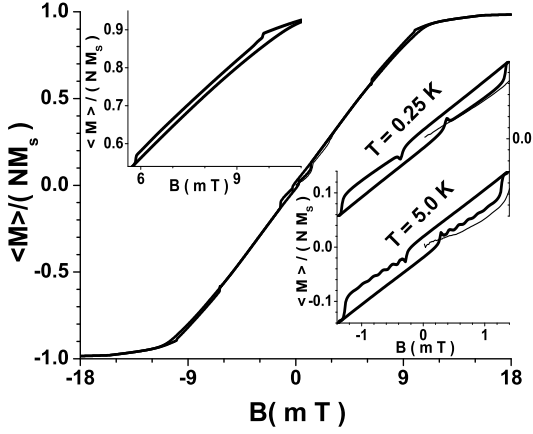}
\caption{The magnetization curves for $N_c=100$ at $T=5$ K,
$a=1.5$ nm, and $\frac{\Delta B}{\Delta t}=3000$ T/s with $B_{\rm
max}=22.5$ mT and $\alpha(T)/\gamma=T_0/T$  for $\hbar g\mu_BB^{
eff}/(k_BT)\ll1$ and $T_0=0.3$ K are shown.\cite{FR} Left inset:
details of the upper portion of the curve.  Right inset:  details
of the central hysteretic portion of the curve shown, along with
the central portion of the corresponding curve at $T=0.25$ K.  The
thin  curves beginning near to the origin represent the
magnetization onsets. These curves are offset for clarity, with
the scales on the right (left) sides corresponding to the lower
(upper) curves, respectively.} \label{fig3}
\end{figure}

 In order to investigate further the differences between the
starting magnetization curves and the curves obtained subsequent
to saturation, in Fig. 4, we have shown the corresponding central
hysteresis loop portion of the magnetization obtained for  two
individual configurations, using the same  experimental and
numerical parameters as in Fig. 3, except that $T=10$ K, for which
$\alpha(T)/\gamma=0.03$. As in Fig. 3, $T$ enters both  through
the statistical averaging and  through the damping, $\alpha(T)$.
In Fig. 4, the solid and open circles correspond to the starting
magnetizations of the two configurations, and the coincident thick
black and thin light grey curves correspond to the central
hysteresis loop region of  their respective magnetization curves
obtained after saturation.  Note that after the initial noisy
regions, the starting magnetizations  for these two configurations
exhibit comparably large amplitude oscillations
 at the frequency $f/2$, the phases of which are very different.
However, after the attainment of the saturation magnetization,
these large amplitude oscillations are absent, and replaced by
smaller amplitude oscillations  at the frequency $f$, which are
similar to the oscillations present in our results obtained at
$T=5$ K shown in the lower curves in the right inset of Fig. 3. We
note that in the first oscillation present on both sides of the
central post-saturation hysteresis loops obtained with these
parameters at $T=5$ and 10 K show additional small amplitude,
higher frequency oscillations,  which may be higher harmonics of
$f$. In addition, the amplitudes of the fifth and sixth
oscillations are larger at 10 K in Fig. 4 than at 5K in the lower
right inset of Fig. 3.

 We remark that the large amplitude oscillations present in the
starting magnetizations shown in Fig. 4 are absent in Fig. 3. This
occurs due to the randomness of the oscillation phases, which is
averaged out in the $N_c=100$ configurations studied in Fig. 2.

 From Fig. 4, we therefore conclude that our starting
configurations that were chosen to have $|M|/M_s\le0.1$,
appropriate for SMM's, lead to starting magnetization curves that
are very different from those that start at the saturation
magnetization, but are subsequently identical.  That is, after the
attainment of saturation, all configurations are identical.

\begin{figure}
\includegraphics[width=0.45\textwidth]{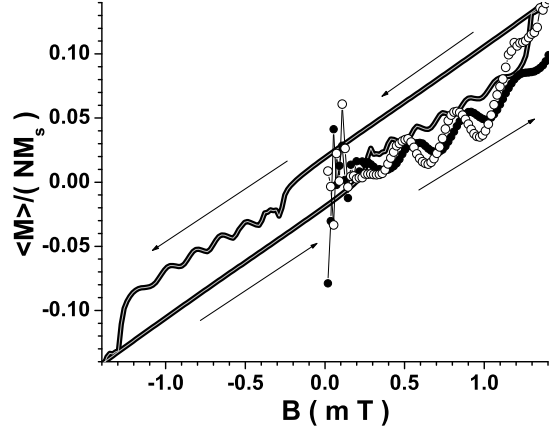}
\caption{The central loop and starting magnetization curves for
two separate configurations, each with $N_c=1$ (open and filled
circles) at $T=10$ K, $a=1.5$ nm, and $\frac{\Delta B}{\Delta
t}=3000$ T/s with $\alpha(T)/\gamma=T_0/T$ for $\hbar g\mu_BB^{\rm
eff}/k_BT\ll1$ and $T_0=0.3$ K are shown.\cite{FR} The thin grey
and thick black curves represent the identical behaviors of the
central hysteretic loop portion of the magnetization for the same
two configurations obtained after saturation. The arrows indicate
the direction of the magnetization hysteresis.  Here $B_{\rm
max}=22.5$ mT.  See text. } \label{fig4}
\end{figure}

\subsubsection{External field directed towards the crystal
corners with $\alpha(T)$}

 We now consider the 3D case of the external magnetic induction
directed from the  crystal center to one of its corners, ${\bm
B}=B(\hat{\bm x}+\hat{\bm y})/\sqrt{2}$, the (110) direction. In
Fig. 5,  we show the resulting central hysteresis region obtained
from our calculations for $N_c=50$, $N=5\times5\times4$, $T=10$ K,
$a=1.5$ nm, and $\frac{\Delta B}{\Delta t}=3000$ T/s with
$\alpha(T)/\gamma=0.03$. In this case, it is sufficient to set
$B_{\rm max}=22.5$ mT, which leads to full saturation.  We note
that for this field direction, a small (-6 mT$ < B < $6 mT)
hysteresis region appears on either side of the origin, which is
rather central to the full magnetization curve, but vanishes over
a small region close to the origin. There are also tiny hysteresis
regions near to saturation that appear as dots in the inset
depicting the full magnetization curve.

The nearly central hysteretic regions shown in Fig.  5 exhibit
reproducible jumps at specific $B$ values, similar to those
observed at low $T$ in SMM's.  However, we note that in this
figure, we have taken $T=10$ K, and have used a classical spin
model.  We also note that we have used a rather small sample
($N=100$) with a fast sweep rate and a large damping coefficient
in our calculations, and caution that such behavior might not be
present in large single crystals, especially with much slower
sweep rates. Nevertheless, this figure demonstrates that steps in
the magnetization do not necessarily have a quantum origin, and
that the sample shape  can lead to unusual hysteresis effects.

\begin{figure}
\includegraphics[width=0.45\textwidth]{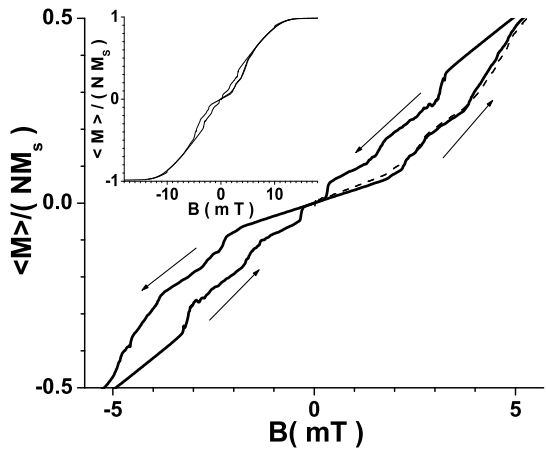}
\caption{The central loops (solid curves) of the magnetization
curve for $N_c=50$, $N=5\times5\times4=100$, with ${\bm B}$ along
the [110] direction [${\bm B}||(\hat{\bf x}+\hat{\bf
y})/\sqrt{2}$] at $T=10$ K, with $a=1.5$ nm, $B_{\rm max}=22.5$
mT, and $\frac{\Delta B}{\Delta t}=3000$ T/s with
$\alpha(T)/\gamma=T_0/T$, and $T_0=0.3$ K. The dashed curve is the
starting magnetization curve.  The arrows indicate the direction
of the hysteresis. Inset: the full magnetization curve.  See
text.} \label{fig5}
\end{figure}

\subsection{Effect of  sweep rate on the hysteresis}

From the curves obtained  using the same numerical parameters as
in Fig. 1 for different induction sweep rates at a fixed, small
damping rate in Fig.  6, it is clear that stronger hysteresis is
found for higher sweep rates, in agreement with experiments on
 a variety of nanomagnets, including SMM's. This shows
that  the reversibility of the process depends on how close to
equilibrium the sweep rate allows the nanomagnet spins to reach.
That is, although for different sweep rates the external induction
is increased by the same amount $\Delta{\bm B}$, at higher sweep
rates,  the time $\Delta t$ allowed for the
 nanomagnets to evolve towards equilibrium is less. This makes the
process less reversible and the hysteresis loops larger.
\begin{figure}
\includegraphics[width=0.45\textwidth]{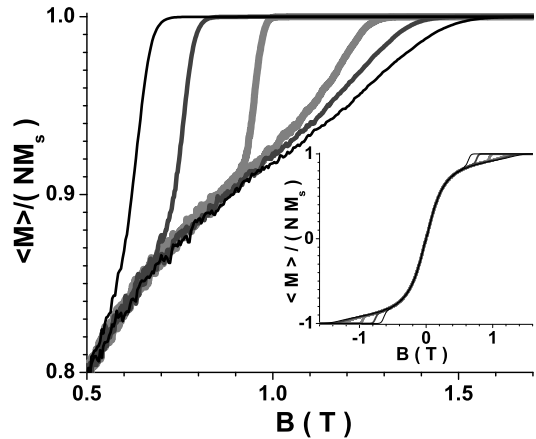}
\caption{Hysteretic region of ${\cal M}(B)$ at 0.7 K,
$\alpha/\gamma=3\times10^{-12}$, and $a$ = 1.5  nm, for  the sweep
rates $\frac{\Delta B}{\Delta t}= 0.04$ T/s (thin black), 0.02 T/s
(dark grey), and 0.005 T/s (thick light grey). The inset shows the
entire curves.} \label{fig6}
\end{figure}

 We also note that at the much higher sweep and damping rates
studied in Figs. 3, 4, the magnetization also exhibits a central
hysteretic region, which exhibits oscillations at  $T$ values not
too low and/or damping constants not too large.

\subsection{Effect of  lattice constant on the hysteresis}

\begin{figure}
\includegraphics[width=0.45\textwidth]{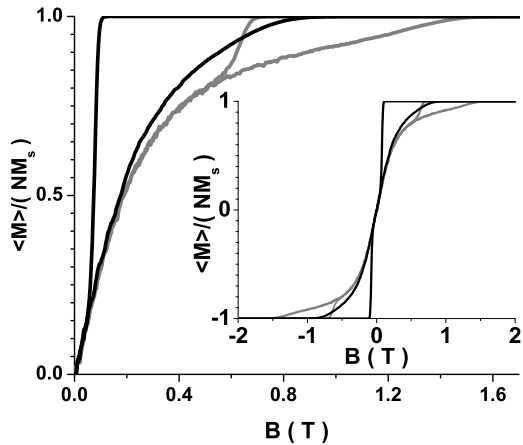}
\caption{Magnetization curves for lattice constants $a$ = 1.5 nm
(grey) and $a$ = 2.5 nm (black). $\frac{\Delta B}{\Delta t} =
0.04$ T/s, $\alpha = 3\times10^{-12}\gamma$, $T$ = 0.7 K. }
\label{fig7}
\end{figure}
 In Fig.  7, we show hysteresis curves for two different values of
the lattice constant $a$,  obtained using the same numerical
parameters as in Fig. 1. For weaker dipole-dipole interactions
(larger $a$), the  rise in the magnetization  is steeper with
increasing ${\bm B}$, and  the rapid decrease in the magnetization
from its saturation value upon decreasing ${\bm B}$ occurs at a
smaller value of $|{\bm B}|$.
  Furthermore, we shall see that dipolar
interactions do not promote hysteresis in these systems, but
suppress it. Actually, the same conclusion was found recently for
magnetic nanoparticles in the framework of the generalized
mean-field
approximation.\cite{meilikhov}\\

This peculiar hysteresis is easily understood by analyzing the LLG
equation.  If the  nanomagnet magnetization ${\bm M}_i^c$ is
parallel to its local magnetic induction
 ${\bm B}_i^{c,\rm eff}$, $d{\bm
M}_i^c/dt=0$,  as it will   remain thereafter, so that ${\bm
M}_i^c$ has reached equilibrium.
  The only chance the system
has to decrease its magnetization  from its saturation value is
through    the combined  weak dipolar  induction, which
strengthens with decreasing lattice parameter $a$.  The dipolar
induction  can oppose the system  from remaining completely
magnetized, since it has small, but non-vanishing components.
 Therefore, even when ${\bm B}$  reaches its maximum (finite) amplitude
 $B_{max}$  and the misalignments of each ${\bm M}_i^c$ with
 ${\bm B}$ are negligible,   dynamic
equilibrium will not generally have been attained due to the
limited time allowed for relaxation before the next change in
${\bm B}$.  There will remain a slight deviation between  the
directions of  the ${\bm B}_i^{c,\rm eff}$ and the ${\bm M}_i^c$
due to the presence of the ${\bm B}_i^{c,\rm dip}$, which is
especially important when $B$ decreases from $B_{\rm max}$.

Of course, it is harder to decrease ${\cal M}$ at the very
beginning of the induction cycle. This is precisely the cause of
the hysteretic behavior, given that changes in ${\bm M}_i^c$ are
proportional to $|{\bm M}_i^c \times{\bm B}_i^{c,\rm eff}|$, which
nearly vanishes when the direction of the incremental induction
has just been reversed. We conclude then that the smaller the
lattice parameter (the stronger the dipolar induction), the
greater the deviation of ${\bm M}_i^c$ from the direction of ${\bm
B}_i^{\rm eff}$. Hence, the easier it is to decrease ${\cal M}$,
making the magnetization curve less hysteretic. This is shown in
Fig.  7, in which the magnetization curves resemble those obtained
for Mn$_4$ SMM's.\cite{wernsdorfer4}  Those data show an abrupt
decrease in ${\cal M}$ at  nearly zero  external induction that is
not  evident in  the magnetization curves of other
SMM's.\cite{vanslageren}

 It is important to note that   the curves in  Figs. 1-7 do
not show the strong hysteresis observed experimentally in  most
SMM's, which is especially large in the central region of the
${\cal M}(B)$  curves.  We remind the reader of our intent to
focus upon the effects of  the dipole-dipole interactions, whereas
the most important  features of SMM's involved in their  low-$T$
relaxation of the magnetization  are generally thought to  be
their quantum structure and magnetic anisotropy. Nevertheless, for
this entirely classical and magnetically isotropic system, we are
indeed finding hysteretic curves.   In addition, the sweep rates
in Figs. 1, 2,  6, and  7 are comparable to those used in
experimental SMM studies. At much larger sweep rates, such as were
studied in Figs.  3-5, an hysteretic central region was found.
However,  the  sizes and $T$ dependencies of these hysteretic
regions were still respectively much smaller and qualitatively
different than observed in SMM's.

\subsection{Effect of spin anisotropy upon the hysteresis}

 It is straightforward to generalize our model to include some of
the effects of magnetic spin anisotropy.  Here we assume the
nanomagnets contain sufficiently many spins that their quantum
nature can be neglected.  We note that SMM's at low $T$ values
behave as quantum entities, because of the small number of spins
in each nanomagnet.  In those systems, most workers have assumed
the in addition to the isotropic Heisenberg and Zeeman
interactions, the magnetic anisotropy terms could also be written
in terms of components of the global spin operator ${\bm S}$, with
the overall dominant terms often written as
$-DS_z^2-E(S_x^2-S_y^2)$.\cite{WS}  However, portions sufficiently
large for model comparison  of the low-$T$ magnetization curves of
two Fe$_2$ SMM dimers   have been studied
experimentally.\cite{shapira1,shapira2} In neither
antiferromagnetic dimer case was any evidence for either of those
types of spin anisotropy present.\cite{ek2}  In contrast,  in one
of those cases, strong evidence for a substantial amount of local,
 single-ion spin anisotropy, in which the individual spins within
a dimer align relative to the dimer axis, is present in the data.
\cite{shapira1,ek2} In addition, the global anisotropy in the
ferromagnetic SMM Mn$_6$ is extremely weak.\cite{morello} Since
the precise quantum nature of more complicated SMM's appears
therefore to be poorly understood, we shall investigate the
quantum features of the magnetic hysteresis curves in SMM's,
including  some effects of local spin anisotropy, in a subsequent
publication.\cite{marisol2}

We therefore restrict our investigations of the role of magnetic
anisotropy upon the magnetization curves of arrays of nanomagnets
to the simplest classical model of spin anisotropy,

\begin{eqnarray}
{\bm B}_i^{c,\rm eff}&=&{\bm B}+{\bm B}^c_{i,\rm dip}+\mu_0{\bm
H}_A,\label{anisotropy}
\end{eqnarray}
where we take  ${\bm B}=B\hat{\bm x}$ and studied the cases ${\bm
H}_A=H_A\hat{\bm x}$ and ${\bm H}_A=H_A\hat{\bm z}$. This is the
3D analogue of the model studied by KS.\cite{saslow} In this
model, the magnetic anisotropy of each of the nanomagnets points
in the same direction, and in our finite sized crystal consisting
of $5\times5\times4$ nanomagnets on a cubic lattice, our chosen
direction is one of the most general ones.  We  first performed
two studies of the magnetic hysteresis in this model, for which
the anisotropy field ${\bm H}_A$ is directed respectively along
(100), $||{\bm B}$,  and (001), $\perp{\bm B}$, and our results
are shown in Figs.  8-9,  respectively. For both  anisotropy field
directions, we take $N_c=100$, $N=5\times5\times4=100$,
$\alpha/\gamma=3\times10^{-12}$, $a=1.5$ nm, $\frac{\Delta
B}{\Delta t}=0.04$ T/s, $T=0.7$ K, and $B_{\rm max}=2.0$ T.   The
sweep rates used in Figs. 8-9 are  slightly faster than those used
in SMM experiments  but much slower than those used in the
calculations of KS. Since $a=1.5$ nm in these curves, these curves
also represent the strongest  realistic dipolar interaction we
studied.

\begin{figure}
\includegraphics[width=0.45\textwidth]{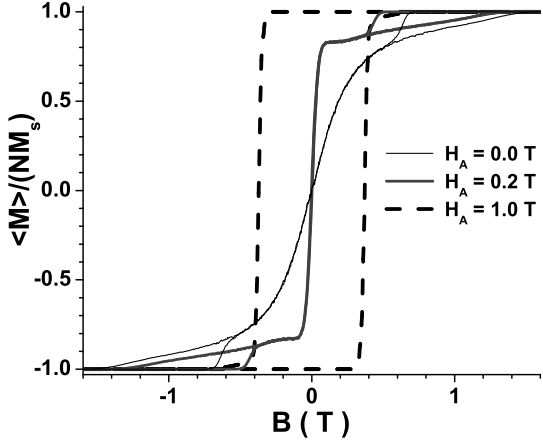}
\caption{Parallel 3D magnetization curves including different
anisotropy field ${\bm H}_A=H_A\hat{\bm x}$ strengths, with  the
external induction ${\bm B}||{\bm H}_A$.  $\mu_0H_A=0$ (thin
black), 0.2 T (dark grey), and 1.0 T (thick dashed), respectively.
For each curve, $N_c=100$, $N=5\times5\times4=100$,
$\alpha/\gamma=3\times10^{-12}$, $a=1.5$ nm, $\frac{\Delta
B}{\Delta t}=0.04$ T/s, $T=0.7$ K, and $B_{\rm max}=2.0$ T.}
\label{fig8}
\end{figure}

 In Fig.  8, we show the portions of the  parallel magnetization
curves with  ${\bm B}||{\bm H}_A||\hat{\bm x}$,  that exhibit the
resulting regions of magnetic hysteresis for three $H_A$ values.
For the $\mu_0H_A=0, 0.2,$ and 1.0 T values shown, all three
curves are hysteretic, but the two lower $H_A$ values do not lead
to a central hysteresis region. Nevertheless, the largest
anisotropy value, $H_A=1.0$ T, leads to a strong central
hysteresis. We remark that the trends shown in Fig. 7 are rather
different from those obtained for a single magnetic particle with
magnetic anisotropy.\cite{fk1}

\begin{figure}
\includegraphics[width=0.45\textwidth]{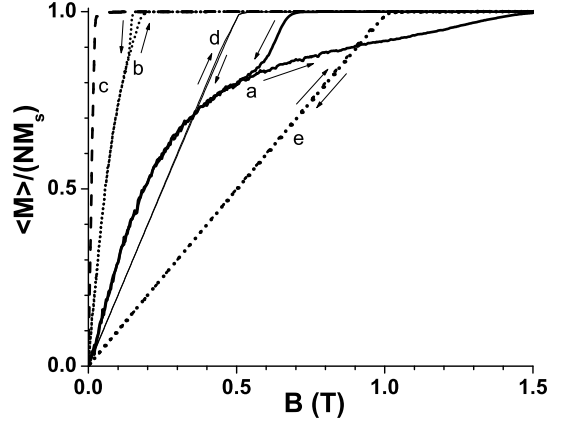}
\caption{Upper region of the 3D perpendicular magnetization curves
with the external induction ${\bm B}=B\hat{\bm x}\perp{\bm
H}_A=H_A\hat{\bm z}$, for different values of $H_A$. Curves
(a)-(e) correspond to $\mu_0H_A=$ 0, $1\times10^{-3}$,
$1.2\times10^{-2}$, 0.5, 1.0 T, respectively. The other parameters
are the same as in Fig. 7. The arrows indicate the directions of
the field sweeps.} \label{fig9}
\end{figure}

 In Fig.  9,
  we show the portions of the  3D perpendicular
magnetization curves exhibiting the resulting regions of magnetic
hysteresis for the five anisotropy fields $\mu_0H_A=0,$ 1 mT, 12
mT, 0.5 T, and 1.0 T, with the magnetic induction ${\bm
B}||\hat{\bm x}\perp{\bm H}_A||\hat{\bm z}$.   In each case,
hysteresis occurs near to magnetic saturation, but is absent in
the central region for small  magnetic induction.  At
$\mu_0H_A=1.0$ T, this is distinctly different from the large
central hysteretic region observed for parallel anisotropy. Note
that  the slope $d{\cal M}/dB$ at small $B$ is non-monotonic with
increasing $H_A$, as it has a minimum at curve (c), corresponding
to $\mu_0H_A=12$ mT.

 Thus, we conclude that it is possible to obtain a central
hysteresis region using this classical model of dipolar
interactions with constant spin anisotropy.  However, our results
suggest that such central hysteresis regions only arise for the
magnetic induction parallel to the spin anisotropy direction, and
for sufficiently strong anisotropy fields, $H_A \ge H_A^{\rm
min}$, where 1.0 T$ > \mu_0H_A^{\rm min}> 0.2$ T.

\section{Dipolar interaction, induction sweep rate, and anisotropy dependencies for a 2D system}

To estimate the importance of the dipolar  induction (especially
when it becomes comparable to the external induction), the
anisotropy and the sweep rate, we have reproduced one of the 2D
calculations of KS.\cite{saslow}  The KS calculation we chose to
reproduce was pictured in their Fig. 2(i), and is shown here as
the left panel of Fig. 10.  Then, we changed some experimental
parameters to see how the results depend on the  anisotropy
strength, sweep rate, and lattice parameter.

Our calculations for a cubic lattice consisting of four
25-particle layers differ from those of KS in many
ways.\cite{saslow} They used a 2D square lattices of cylindrical
nanodots (here, we take their $5\times5$ lattice with external
induction aligned along an array's diagonal), included a
shape-dependent anisotropy field perpendicular to the lattice,
performed their calculations at $T = 0$, used a much larger
damping constant than we  generally did {it for 3D systems}, and
did not average their results over an ensemble  of 2D samples,
because such systems do not show variations in the resulting
hysteresis loops for different initial states. Nevertheless, we
both integrated the LLG equation using the Runge-Kutta algorithm,
and surprisingly, KS's system turned out to be very sensitive to
the dipolar field strength. The effective induction they
considered can be written as
\begin{eqnarray}
{\bm B}_i^{c,\rm eff}&=&{\bm B}+{\bm B}^c_{i,\rm
dip}+\mu_0H_A\hat{\bm z}.
\end{eqnarray}
For lattice constant $a = 1.5$ nm,  spin $S = 5$,  and
$V/a^3=0.5$, where $V$ is the volume of the nanomagnet, the
saturation magnetization is $M_s\approx55$ Oe. Then, they took the
dimensionless $dt = 5 \times 10^{-3}$, which implies a real  time
interval $dt = 5.17 \times 10^{-12}$ s. If the system evolves
during 700 time steps $dt$ for some fixed value of ${\bm B}$, then
${\bm B}$ is changed every $\Delta t\approx 3.62 \times 10^{-9}$
s. On the other hand,  KS  chose a maximum external induction
${B}_{\rm max} = 2\mu_0 M_s \approx 1.1 \times 10^{-2}$ T.  In
addition, they took fixed induction steps of  magnitude $\Delta B
= 2\times10^{-3}\mu_0 M_s \approx 1.1 \times 10^{-5}$ T.
Therefore, we estimate  their resulting sweep rate  to be
$\frac{\Delta B}{\Delta t}\approx3\times10^3$ T/s,  as in our 3D
results shown in Figs. 3, 4.

 In the absence of any specific information, we then had to
induce the value of the anisotropy field  that KS used  to obtain
their figure. Fortunately, as discussed in the following, the
results are rather insensitive to it, unless $H_A$ becomes
comparable to $B_{\rm max}/\mu_0$.  In the right panel of Fig. 10,
our 2D calculation with $\mu_0H_A=0.75$ mT are shown, and by
comparing that figure with Fig. 2(i) of KS pictured in the left
panel of Fig. 10, we see that the agreement is remarkably good.
\begin{figure}
\vspace*{20pt}{\includegraphics[width=0.23\textwidth,height=0.23\textwidth]{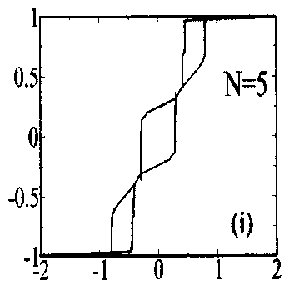}}\hspace*{2pt}
{\vspace*{-10pt}\includegraphics[width=0.23\textwidth,height=0.23\textwidth]{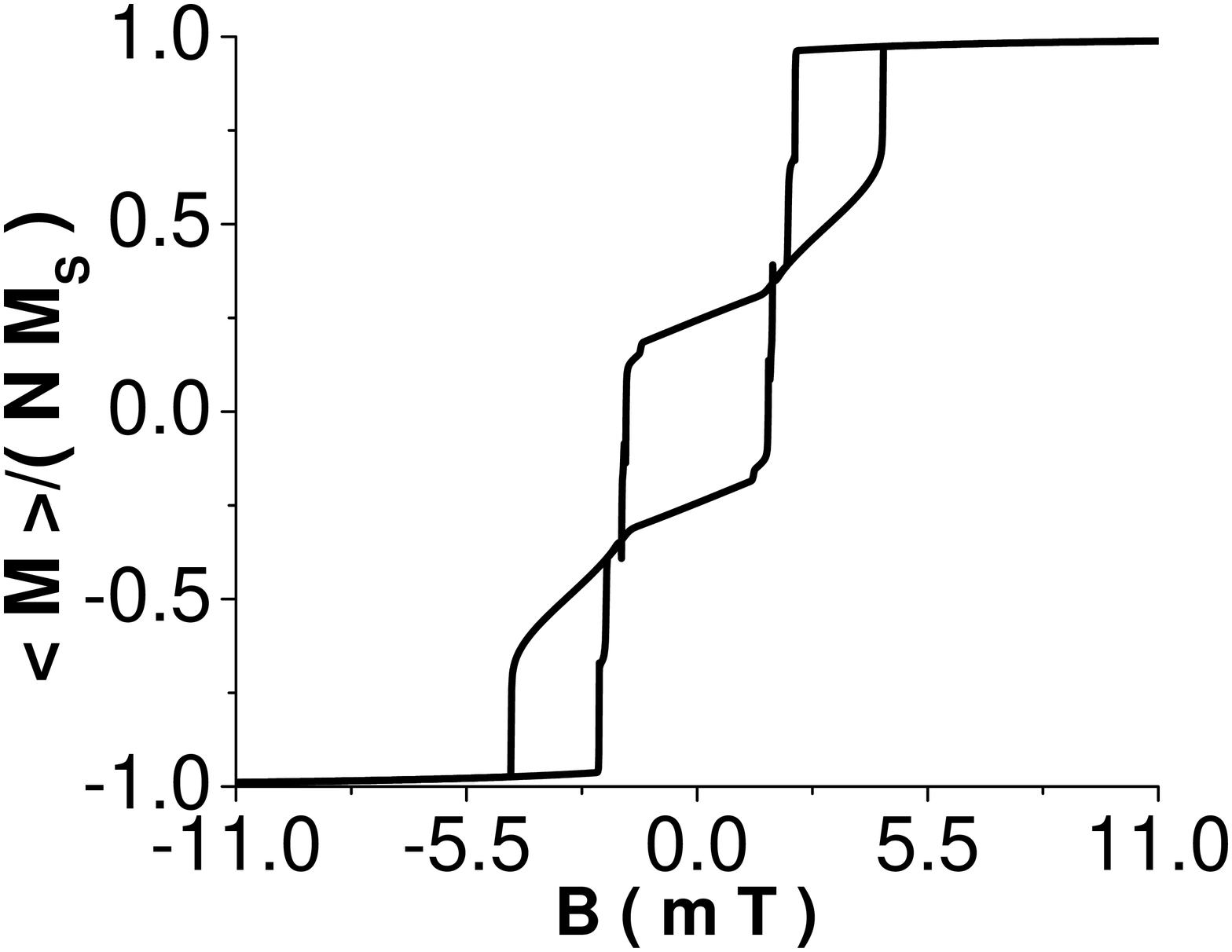}}
\vspace*{10pt} \caption{(left) Hysteresis loop ${\cal M}(B)$ in
units of $M_s$, for a weakly coupled array of $5\times5$
ferromagnetic nanodots in a square lattice on the $xy$ plane, from
Fig. 2(i) of KS. The external induction is applied along the array
diagonal ($45^{\circ}$ from the $x$ axis).\cite{saslow}. (right)
Our results calculated for $N_c=1$ with $5\times5$ nanomagnets on
a square lattice, $\alpha/\gamma = 0.6$, $T = 0$K, $\frac{\Delta
B}{\Delta t} = 3000$ T/s, $\mu_0H_A = 7.5\times10^{-4}$ T, ${\bm
B} = B(\hat{\bm x}+\hat{\bm y})/\sqrt{2}$.} \label{fig10}
\end{figure}

Very recently, Takagaki and Ploog (TP) used a fourth-order
Runge-Kutta procedure to integrate the LLG equations with $N\times
N$ 2D nanomagnet lattices with magnetic anisotropy and
dipole-dipole interactions.\cite{tp} They used a fixed time
interval $dt=0.1\hbar/(\gamma M_s)$, 20 times as fast as that used
by KS,\cite{saslow} and continued interating until no further
changes in the nanomagnet spin configurations were obtained.  They
obtained results for $N=5$ which they described as considerably
different from those of KS, with a somewhat different
magnetization loop and a larger area of the hysteretic
regions.\cite{tp} Although they claimed that their fourth-order
procedure was intrinsically more accurate than the second-order
one used by KS and by us, the fact that we obtained the excellent
agreement pictured in Fig. 10 with one of the $N=5$ results of KS
suggests that the  procedure used by TP  might have been less
accurate than they claimed.\cite{tp}

\subsection{Anisotropy field dependence of the hysteresis}

\begin{figure}
\includegraphics[width=0.45\textwidth]{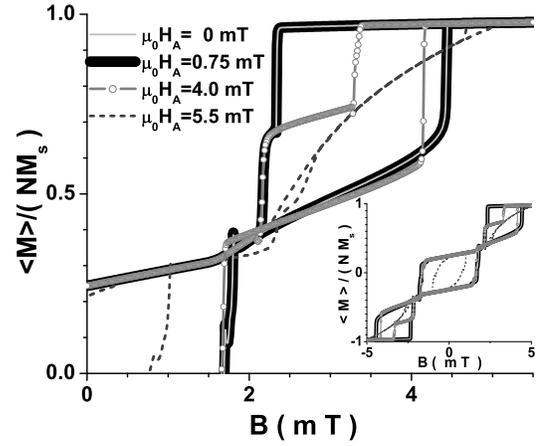}
\caption{Hysteresis loops for different strengths of  $H_A$ for
  $5\times5$ nanomagnets on a square  lattice with $N_c=1$. $S = 5$, $T = 0$ K, $a =
1.5$ nm, $\frac{\Delta B}{\Delta t} = 3000$ T/s, $\alpha/\gamma =
0.6$, ${\bm B} = B(\hat{\bm x}+\hat{\bm y})/\sqrt{2}$. The thin
grey and thick black curves with
 $\mu_0H_A = 0,
0.75$ mT, respectively, are nearly indistinguishable. The small
grey circles and dashed curves correspond to $\mu_0H_A=4.0, 5.5$
mT, respectively.  The inset shows the entire curves, which are
symmetric with respect to the origin. }\label{fig11}
\end{figure}

 We first investigated the effects of changing the strength of
the anisotropy fields, and presented our results in Fig. 11. The
most important issue about the results shown in Fig. 11 is the
fact that the curve  obtained by KS  (the left panel of Fig. 10)
is basically independent of the anisotropy field $H_A$  for
sufficiently small $H_A$.   That is, there are no essential
differences between that curve  reproduced in the right panel of
Fig. 10 with $\mu_0H_A=7.5$ mT, and the one with $H_A=0$. Strong
deviations from these essentially identical curves appear for
$\mu_0H_A\ge 4$ mT, however.   Since identical behavior is
obtained without any anisotropy, this implies
 that all hysteretic features (including the stepped
magnetization and demagnetization) are due to the dipolar
interaction. $H_A$ becomes important only when it is comparable to
$B_{\rm max}/\mu_0$ and tends to close the hysteresis loops,
starting from the lower and upper loops.

 We note that by comparing Fig. 11 with Fig. 9, the details of
the hysteresis obtained with $H_A=0$ for ${\bm B}$ along the (110)
direction are different in 3D and 2D samples.  The hysteresis is
much larger in the 2D case pictured in Fig. 11, and has a large
loop in the central region that does not vanish at the origin,
plus large loops that extend up to saturation.  In the 3D case
constructed from four 2D planes each equivalent to that used in
the calculation shown in Fig. 11, the magnitude of the hysteresis
is reduced and its details have been greatly altered.

\subsection{Induction sweep rate dependence of the hysteresis}

  In Fig. 12, we show our  results for a single configuration of a square 2D lattice with $N=5\times5$ for different sweep rates, keeping
 the other parameters fixed at $\mu_0H_A=0.75$ mT, $\alpha/\gamma=0.6$, $a=1.5$ nm, $S=5$,
$T=0$, and ${\bm B}=B(\hat{\bm x}+\hat{\bm y})/\sqrt{2}$. From
Fig. 12,  we note  that the hysteresis is nearly independent of
induction
 sweep rate  over the range 300 to 6000 T/s,  distinctly different from
 the strong dependence found in 3D systems  shown in Fig. 5.

\begin{figure}
\includegraphics[width=0.45\textwidth]{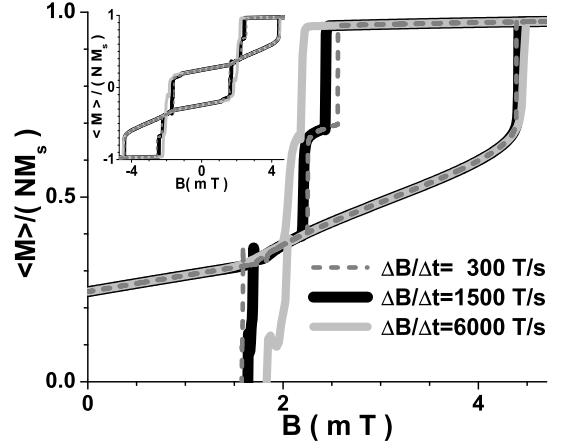}
\caption{Hysteresis loops for different induction sweep rates with
 $5\times5$ nanomagnets on a square
lattice with $N_c=1$. $S = 5$, $T = 0$ K, $a = 1.5$ nm, $\mu_0H_A=
0.75$ mT, $\alpha/\gamma = 0.6$. The dashed grey, thick black, and
light solid grey curves correspond to $\frac{\Delta B}{\Delta
t}=300, 1500, 6000$ T/s, respectively.  The inset shows the entire
curves. ${\bm B} = B(\hat{\bm x}+\hat{\bm
y})/\sqrt{2}$.}\label{fig12}
\end{figure}

\subsection{Lattice parameter dependence of the hysteresis}

 In Fig. 13, we have illustrated the effect of the lattice
constant $a$ upon the hysteresis.  In this figure, we  kept the
other parameters fixed at $S=5$, $T=0$, $\frac{\Delta B}{\Delta
t}=3000$ T/s, $\mu_0H_A=0.75$ mT, $\alpha/\gamma=0.6$, and ${\bm
B}=B(\hat{\bm x}+\hat{\bm y})/\sqrt{2}$. As  $a$ is varied from
2.0 to 1.25 nm, the upper portions of the hysteresis curves appear
from left to right, respectively. From Fig. 13, it is readily seen
that the magnetization curves are very sensitive to $a$ and hence
to the strength of the dipolar interaction, which is proportional
to $a^{-3}$. Our results for $a$=2.5 nm exhibit a smaller
hysteresis shifted further to the left, and all indications of
steps have disappeared. Although  not shown  in Fig. 13, as $a$ is
increased further to 3.0 nm, the hysteresis almost disappears
entirely. We  deduce that stronger dipolar interactions  (smaller
$a$) result in larger hysteresis loops containing increased widths
of additional steps.

\begin{figure}
\includegraphics[width=0.45\textwidth]{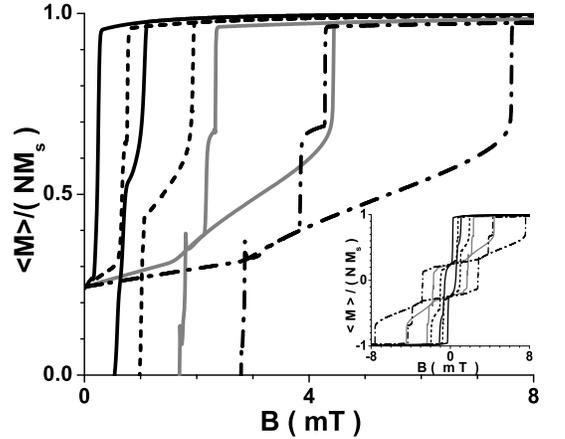}
\caption{Hysteresis loops for lattice parameters $a = 2.5$ nm
(solid black), $a = 2.0$ nm (dashed black), $a = 1.5$ nm (solid
grey), and $a = 1.25$ nm (dot-dashed black), for  $5\times5$
nanomagnets on a square lattice with $N_c=1$. $S = 5$, $T = 0$ K,
$\frac{\Delta B}{\Delta t} = 3000$ T/s, $\mu_0H_A=
7.5\times10^{-4}$ T, $\alpha/\gamma = 0.6$. The inset shows the
entire curves. ${\bm B} = B(\hat{\bm x}+\hat{\bm
y})/\sqrt{2}$.}\label{fig13}
\end{figure}

 We then  infer that contrary to the
conclusion found for the 3D systems (based upon  much smaller
damping coefficients and much slower sweep rates), the dipolar
interactions promote a hysteretic behavior in this 2D system.

\section{Summary and Conclusions}

  We first found $N_c=100$ sample configurations with an overall
magnetization close to 0. We then solved the
Landau-Lifshitz-Gilbert equation for a 3D cubic lattice of
$N=5\times5\times4$ nanomagnets, subject to dipole-dipole
interactions and spin anisotropy.  These results should be
relevant for an array of Stoner-Wolfarth nanomagnets, and to some
extent, single molecule magnets, although the quantum nature of
the latter has so far been neglected.  In the absence of spin
anisotropy, we varied the magnetic induction sweep rate
$\frac{\Delta B}{\Delta t}$, the damping constant $\alpha$, the
lattice constant $a$, and the temperature $T$.  We also considered
the effects of a $T$-dependent damping constant of the form
$\alpha(T)/\gamma=T_0/T$ suggested by Fredkin and Ron. For slow
sweep rates and small $\alpha$ relevant for experimental studies
on single molecule magnets, magnetic hysteresis appears in the
regions of the magnetization curves near to saturation, the area
and onset magnetic induction strength of which increases with
decreasing $\alpha$ and increasing sweep rate.  With decreasing
$T$, the onset magnetization magnitude of the hysteretic regions
near to saturation decreases.  With decreasing $a$ corresponding
to increased dipole-dipole interaction strengths, the onset of the
hysteresis regions near to saturation appears at increasing
magnetic induction magnitude.

At much larger sweep rates and damping constants, the
magnetization curves attain saturation at much smaller applied
magnetic induction strengths.  The hysteretic regions just below
saturation have moved somewhat below saturation, and a new central
hysteretic region appears.  As one follows the magnetization curve
for a single configuration, the starting curve exhibits
oscillations at a rather constant (magnetic induction independent)
frequency $f/2$, but the phase of the magnetization oscillations
is a random function of the configuration.  After the attainment
of magnetic saturation, this central hysteretic region exhibits
oscillations at $f$, twice that frequency,   possibly with weak
higher harmonics, for $T$ not too low, which are independent of
the configuration.

When the applied magnetic induction is in the (110) direction
(from the sample center to one of its corners), magnetic
hysteresis exhibiting steps and jumps appears within the central
region, but vanishes at and very near to the origin.  Although
these step-like features are suggestive of the behavior seen in
single molecule magnets, they are present at rather high $T$
values, as they arise from the classical sample shape effects.

In the presence of the magnetic anisotropy field ${\bm H}_A$, an
applied magnetic induction parallel to the anisotropy axis leads
to a large central hysteresis region, provided that the magnitude
of the spin anisotropy is sufficiently large.  For the applied
magnetic induction perpendicular to the magnetic anisotropy, no
central hysteresis region is present, although a small amount of
hysteresis near to saturation persists for sufficiently small spin
anisotropy strength, and the slope of the magnetization curve at
the origin is non-monotonic, exhibiting a maximum at a particular
small value of the spin anisotropy strength.  These effects for
the spin anisotropy are qualitatively in agreement with those in
many types of arrays of nanomagnets, including single molecule
magnets.

As a test of our numerical procedure, we studied the simplified
$5\times5$ 2D square lattice with a perpendicular spin anisotropy
field ${\bm H}_A$ using the same procedure, and for a particular
set of parameters, obtained quantitative agreement   with a
hysteresis curve obtained for that system by Kayali and
Saslow.\cite{saslow} We showed that their hysteresis curve is
basically independent of  $H_A$  until $\mu_0H_A$  is on the order
of the external induction. We also  demonstrated that the magnetic
hysteresis does not depend significantly upon the magnetic
induction sweep rate,  as opposed to the  dependence we found in
our 3D system.  In addition, we found  that the magnetization of
the 2D system is very sensitive to variations in the lattice
parameter $a$. Finally, we noticed that although dipolar
interactions  also oppose the magnetization process in 2D systems,
 increasing the onset magnetic induction strength for the
attainment of saturation as in 3D systems, they increase the area
of the hysteresis, a behavior opposite to that found for the 3D
system with a much smaller damping coefficient and much slower
sweep rate.

 We expect our results to be relevant to the magnetization
processes in a variety of nanomagnet arrays, especially those
approximating arrays of Stoner-Wolfarth particles.  In addition,
some of the features we obtained should be relevant to single
molecule magnets, although the temperature dependence of the
effects is not in agreement with experiments on those materials.
Further studies of the magnetic hysteresis using quantum models of
the nanomagnets and their various anisotropy types is warranted,
and will be addressed subsequently.\cite{marisol2}

\section{Acknowledgments}
This work was supported in part by the NSF  under grant number
NER-0304665.

\section{Appendix A}

We rotate our reference frame at every integration time step in
such a way that ${\bm B}_i^{c,\rm eff}$ is along the $z$ axis. In
this case, we can easily solve the LLG equation, Eq. (1). For
simplicity of notation, we drop the subscripts $i$ and
superscripts $c$, and remember that we are describing the
precession of the $i$th nanomagnet in the $c$th crystal.  We
define the axes to describe the magnetization direction of this
particular nanomagnet, $\hat{\bm M}$, $\hat{\bm \theta}$, and
$\hat{\bm\phi}$, where $\hat{\bm\phi}=\hat{\bm
M}\times\hat{\bm\theta}$, and then write
\begin{eqnarray}
{\bm B}^{\rm eff}&=&B_z\hat{\bm z}=B_z(\hat{\bm
M}\cos\theta-\hat{\bm\theta}\sin\theta)\nonumber\\
&=&\hat{\bm M}B_M+\hat{\bm\theta}B_{\theta}.
\end{eqnarray}
 Since the magnitude of the
dipole moment $M_s$ is conserved, in spherical coordinates Eq.(1)
leads to
\begin{eqnarray}
\frac{d\hat{\bm
M}}{dt}&=&\hat{\bm\theta}\frac{d\theta}{dt}+\hat{\bm\phi}\sin\theta\frac{d\phi}{dt}\nonumber\\
&=&\hat{\bm\theta}\alpha B_{\theta}+\hat{\bm\phi}\gamma
B_{\theta}.
\end{eqnarray}

Finally, from \begin{eqnarray} \frac{d\theta}{dt}&=&=-\alpha|{\bm
B}^{\rm eff}|\sin\theta,\label{thetadot}\\
\frac{d\phi}{dt}&=&-\gamma|{\bm B}^{\rm eff}|,\label{phidot}
\end{eqnarray} we obtain  for a very small time interval $dt$,
\begin{eqnarray}
\phi(t_0+dt)&\approx&\phi(t_0)-\gamma|{\bm B}^{\rm eff}(t_0)|dt,\label{phioft}\\
\theta(t_0+dt)&\approx&\theta(t_0)-\alpha|{\bm B}^{\rm
eff}(t_0)|\sin[\theta(t_0)]dt.\label{thetaoft}
\end{eqnarray}
These equations were used in our numerical calculations.  In order
to relate the angles to measurable quantities, however, we note
that it is possible to integrate Eqs. (\ref{thetadot}) and
(\ref{phidot}) exactly, obtaining
\begin{eqnarray}
\theta(t)&=&\cos^{-1}\Bigl[\tanh\Bigl(\tanh^{-1}(\cos[\theta(t_0)]\nonumber\\
& &+\alpha\int_{t_0}^td\tau|{\bm
B}^{\rm eff}(\tau)|\Bigr)\Bigr],\label{theta}\\
\phi(t)&=&\phi(t_0)-\gamma\int_{t_0}^td\tau|{\bm B}^{\rm
eff}(\tau)|,\label{phi}
\end{eqnarray}
which is equivalent to that obtained using a somewhat different
technique.\cite{chen}  We note that by expanding Eqs.
(\ref{theta}) and {\ref{phi}) to leading order in $dt$, we recover
Eqs. (\ref{thetaoft}) and (\ref{phioft}), respectively.

However, these more general forms for $\theta(t)$ and $\phi(t)$
lead to a more physical interpretation of our method. Since the
dimensionless magnetization components along and perpendicular to
${\bm B}^{\rm eff}$ are $M_z=\cos\theta$,
$M_x=\sin\theta\cos\phi$, and $M_y=\sin\theta\sin\phi$, we have
\begin{eqnarray}
M_{z}(t)&=&\tanh\Bigl(\tanh^{-1}[M_{z}(t_0)]+\alpha\int_{t_0}^td\tau|{\bm
B}^{\rm eff}(\tau)|\Bigr),\nonumber\label{Mzoft}\\ & &\\
M_x(t)&=&\sqrt{1-[M_{z}(t)]^2}\cos[\phi(t)],\label{Mpoft}\\
\noalign{and}\nonumber\\
 M_y(t)&=&\sqrt{1-[M_{z}(t)]^2}\sin[\phi(t)].\label{Mpboft}
\end{eqnarray}

 Independent of the coordinates, we must assure that (for the
$i$th nanomagnet in the $c$th configuration) ${\bm M}$ changes its
direction smoothly, in order to obtain a reliable calculation for
the overall $\overrightarrow{\cal M}$. Since each component of
${\bm M}$ cannot change dramatically, we must therefore require
$\theta\ll2\pi$ and $\phi\ll2\pi$.  These restrictions then
require us to set the time integration step width $dt$
sufficiently small.  If for example, $\gamma/\alpha$ were on the
order of $10^{+11}$ and $|{\bm B}^{\rm eff}|$ were in the range
$10^{-3}-10^{-2}$ T, we would require $dt<10^{-11}$ s.  For sweep
rate $\frac{\Delta B}{\Delta t}\approx10^{-2}$T/s, where $\Delta
t=N_tdt\approx10^{-4}$ s, $N_t$ must be on the order of $10^7$.
Since we would need to recalculate the direction of the
magnetization of each nanomagnet $N_t$ times in each $\Delta B$
step, this would be a significant challenge with present day
computers.

 One thing we can do to make  our calculations feasible for the
sweep rates used in SMM studies is to set $\alpha$ extremely
small, say $\alpha/\gamma\alt 10^{-10}$,  although such small
$\alpha$ values have not been reported in experiments. Otherwise,
to study much larger but perhaps more reasonable $\alpha$ values,
we would have to use much faster sweep rates, as in
KS.\cite{saslow}

\end{document}